\documentclass[12pt]{article}
\usepackage{geometry}
\usepackage[applemac]{inputenc}
\usepackage{amsmath}
\usepackage{graphicx}
\usepackage[colorlinks]{hyperref}

\title{Unambiguous Quantum Gravity Phenomenology Respecting Lorentz Symmetry}

\author{Yuri Bonder \\ yuri.bonder@nucleares.unam.mx \\[2ex]
Daniel Sudarsky \\ sudarsky@nucleares.unam.mx\\ \\ Instituto de Ciencias Nucleares\\
Universidad Nacional Aut\'{o}noma de M\'{e}xico
\\ A. Postal 70-543, M\'exico D.F. 04510, M\'exico}

\begin{document}

\maketitle

\begin{abstract}
{We describe a refined version of a previous proposal for the exploration of quantum gravity phenomenology. Unlike the original scheme, the one presented here is free from sign ambiguities while it shares with the previous one the essential features. It focuses on effects that could be thought as arising from a fundamental granularity of quantum space-time. The sort of schemes we consider are in sharp contrast with the simplest scenarios in that such granularity is assumed to respect Lorentz Invariance but it remains otherwise unspecified. The proposal is fully observer covariant, it involves non-trivial couplings of curvature to matter fields and leads to a well defined phenomenology. We present the effective Hamiltonian which could be used to analyze concrete experimental situations, and we shortly review the degree to which this proposal is in line with the fundamental ideas behind the equivalence principle.}
\end{abstract}


\section{Introduction}

The idea of accessing empirically some aspects of quantum gravity has been receiving a great deal of attention during the last few years. The main directions of research have focussed on possible violations, or deformations, of the space-time symmetries, particularly Lorentz invariance. The motivation for that kind of considerations was essentially the following: Any sort of space-time granularity, presumably associated with the Planck scale, should be incompatible with the Lorentz length contraction of special relativity, and thus, quantum gravity would naturally lead to either modifications or violations of Lorentz invariance. This drew the attention of the quantum gravity community on the general subject of phenomenology of Lorentz invariance violation, rekindling interest on a much older program \cite{Initial-LIV, SME}. The most direct approach considers invoking an actual violation of special relativity, associated with the existence of a preferential rest frame on which the granularity takes its most symmetric form, and most naturally identified with the local frame singled out by the cosmological co-moving observers (operationally, the local frame where the CMB dipole vanishes). In this context, relatively detailed theoretical proposals where made based on String Theory \cite{ST} and Loop Quantum Gravity \cite{LQG}. Moreover, a substantial program looking for direct manifestations of these effects has lead to remarkable bounds that essentially rule out the effects which are not suppressed at the level of energies of about a billion times larger than the Planck Scale \cite{Bounds}. Furthermore, when these ideas are taken at face value and combined with simple quantum field theoretical calculations of the radiative corrections, one finds that the natural size of the effects is not suppressed by the ratio of the characteristic energies to the Planck Energy, but merely by the standard model coupling constants \cite{Collins}. This has led us 
to the conclusion that, if such granular structure of space-time associated with a preferential reference frame were real, the effects would have been noticed long ago, and thus that what Nature is telling us is that Lorentz invariance is not broken by any such granular structure of space-time.

Of course, one should bear in mind that a space-time granularity does not by itself imply a violation of Lorentz invariance \cite{Sorkin-Rovelli}. Moreover, it is not at all clear that space-time should be granular at any scale, and if we consider it to be granular at some scale, we still need to answer what would it mean to be granular at a certain scale if we can not point out the frame at which any hypothetical scale can be said to characterize the fabric of space-time. We should mention that at this point there is no good geometrical picture of how a granularity might be associated to space-time while strictly preserving the Lorentz and Poincar\'e symmetries (the only scheme we know that can be reasonably expected to lead to any such detailed picture is the Poset program \cite{Poset}, which is however very much under development). We must say that our current views of what is quantum gravity, or even what would be the appropriate language to talk about a quantum space-time that is probed by fundamentally quantum objects, are likely to turn out to be extremely naive \cite{Unspekables}, thus, in developing this proposal we rely heavily on simple analogies combined with symmetry considerations, which we consider to be a rather conservative approach given that at this time we do not have a fully satisfactory theory of quantum gravity.

Leaving aside such issues, in this line of work we are motivated by the following relatively simple question: If space-time has some sort of fundamental granularity which respects Lorentz symmetry, how could it possibly become manifest? In a previous work \cite{NewQGP} a proposal was made in which a granular structure of space-time might become manifest in a rather subtle way so that it would be immune from the previous considerations while still, in principle, susceptible to a phenomenological study. That proposal had a couple of problematic aspects and some of them are discussed in \cite{QGPwLSV} where a refined proposal was given.

The guiding analogy behind the original proposal is the following: One starts by considering a hypothetical experimental physicist from another planet who is trying to uncover the granular structure of a salt crystal. He believes that the fundamental symmetry of this crystal is anything but cubic and he has chosen to carry out his investigations using a macroscopically cubic piece of crystal. The researcher is hoping to uncover evidence of the fundamental granularity by looking for experimental signals that indicate deviations from what he considers to be just the macroscopic symmetry of his crystal. We know that he will find none simply because the fundamental structure of the crystal is also the cubic symmetry. This, we believe, is the situation we face today regarding the fundamental granular structure of space-time (assuming that such granularity exists) and the attempts to seek evidence for its existence through deviations from strict Lorentz invariance. As in the case of the crystal, the discrete structure might be studied, but not by looking at deviations from the underlying fundamental symmetry. The point is that when one considers a macroscopic crystal whose global form is not compatible with the structure of the fundamental crystal, say spherical, the surface will necessarily include some roughness, and thus, a manifestation of the granular structure will occur through the breakdown of the exact spherical symmetry.

Taking this as a guiding lesson and sensibly extrapolating to the case of the micro and macro structures of space-time, we are driven to start with the assumption that the symmetry of the fundamental building blocks of space-time is the Lorentz symmetry, and thus, that we would expect no violation of this symmetry at the macroscopic level to the degree to which the space-time is macroscopically Lorentz invariant on an extended domain. That is, in a region of space-time which could be considered as well approximated by the Minkowski metric, the granular structure of the quantum space-time would not become manifest through the breakdown of its symmetry. However, and following our solid state analogy, we are led to consider the situation in which the macroscopic space-time geometry is not fully matched with the symmetry of its basic constituents. The main point is that in the event of a failure of the space-time to be exactly Minkowski in an extended region, the underlying granular structure of quantum gravity could possibly become manifest affecting the propagation of the various matter fields. Such situation should thus be characterized in our macroscopic description of space-time by the Riemann tensor, which is known to describe the failure of a space-time to be Minkowski in an open region. Therefore, the non-vanishing of the Riemann tensor would characterize situations where the non-trivial aspects of the microscopic structure of space-time might become manifest. Moreover, due to the implicit correspondence of the macroscopic description with the more fundamental one, we can expect that the Riemann tensor would also indicate the space-time directions associated with the sought effects. We should note that, in contrast with many other proposals of quantum gravity phenomenology, the type of schemes we are dealing with do not invoke any notion of a global preferential frame\footnote{Some authors use this notion in a `weak sense' taking it to refer to the absence of one single special frame selected by all the new objects of the theory (e.g., \cite{SME2}), but we refer here to the `strong sense' of the notion, \textit{i.e.}, that there should be no way to use any new object of the theory to identify any spacial space-time direction or set of directions.}, nor does it call for new fields (be them dynamical or non-dynamical), and in fact, by construction, it leads to absolutely no new effects or deviations from standard physics in any perfectly Minkowskian region of space-time.

Based on the general considerations above, we are looking for an effective description of the ways in which the Riemannian curvature could affect, in a nontrivial manner, the propagation of matter fields. The effective description of such a thing should involve Lagrangian terms representing the coupling of matter fields with the Riemann tensor. However, the Ricci tensor represents that part of the Riemann tensor which, at least on shell, is locally determined by the energy-momentum of matter fields at the events of interest. Thus, any coupling of matter fields to the Ricci tensor would, at the phenomenological level, reflect a sort of point-wise self interaction of matter that would amount to a locally defined renormalization of the usual phenomenological terms such as a the mass or the kinetic terms in the Lagrangian. But we are interested in the underlying structure of space-time rather than the self interaction of matter, therefore, we need to ignore the aspects that encode the latter, which in our case corresponds to all Lagrangian terms containing the Ricci tensor coupled to matter fields. The remainder of the Riemann tensor, \textit{i.e.}, the Weyl tensor, can thus be thought to reflect the aspects of the local structure of space-time associated solely with the gravitational degrees of freedom.

Note that, in the absence of gravitational waves, the Weyl tensor is connected with the nearby `matter sources' rather than just the matter present in the point of interest, and moreover, that such connection involves the propagation of the influence of such sources through the space-time and thus, the structure of the latter would be playing a central role in the way the influences might become manifest. In this sense the Weyl tensor reflects the `non-local effects' of the matter in contrast with the Ricci tensor which is fully determined point-wise by the matter fields. The task is then to consider non-trivial ways to couple the Weyl tensor to matter fields in the standard model of particle physics. We will focus here on the fermionic fields $\Psi$.

\section{The previous proposal} \label{Previous}

Let us briefly describe the model presented in \cite{QGPwLSV} in order to see the ambiguity we will address in the present work. In that case the goal was to construct a Lagrangian term where the fermionic fields are coupled to the Weyl tensor\footnote{We are using Greek letters as space-time indexes, Latin letters represent spacial indexes. Riemann normal coordinates about the point of interest are a used.} $W_{\mu\nu\rho\sigma}$ and which is `minimally suppressed' by the Planck mass $M_{Pl}$. The most obvious term one can write is $W_{\mu\nu\rho\sigma} \bar{\Psi}\gamma^\mu \gamma^\nu \gamma^\rho \gamma^\sigma \Psi$, however, this term vanishes. Therefore we must seek terms that are either highly suppressed or less natural. Following \cite{NewQGP} we take a particular approach of the latter type.

We consider the $(2,2)$ Weyl tensor ${W_{\mu\nu}}^{\rho\sigma}$ as a mapping from the space of $2$-forms $\mathcal{S}$ into itself. As is well known, the space-time metric endows the six dimensional vector space $\mathcal{S}$ with a pseudo-Riemannian `supermetric' $\mathcal{G}$. With the aid of the Weyl tensor we constructed two self-adjoint operators on $({\mathcal{S}},\mathcal{G})$:
\begin{eqnarray}
\label{W+}{{(W_+)}_{\mu\nu}}^{\rho\sigma}&=&\frac{1}{2}\left({W_{\mu\nu}}^{\rho\sigma}+{W^\dagger_{\mu\nu}}^{\rho\sigma}\right),\\
\label{W-}{{(W_-)}_{\mu\nu}}^{\rho\sigma}&=&\frac{i}{2}\left({W_{\mu\nu}}^{\rho\sigma}-{W^\dagger_{\mu\nu}}^{\rho\sigma}\right),
\end{eqnarray}
where ${W^\dagger_{\mu\nu}}^{\rho\sigma}={W^{\rho\sigma}}_{\mu\nu}$. The operators given in equations (\ref{W+}) and (\ref{W-}) can be diagonalized having a complete set of eigenforms $\Xi^{(\pm,s)}_{\mu \nu}$, where the sign in the super-index specifies to which operator the eigenforms correspond while $s$ is an index running from $1$ to $6$. The original idea was to use the normalized non-null eigenforms and the non-vanishing eigenvalues, $\lambda^{(\pm,l)}$, of ${{(W_\pm)}_{\mu\nu}}^{\rho\sigma}$ to construct an object that can be coupled to $\bar{\Psi}\gamma_{[\mu} \gamma_{\nu]} \Psi$.

The first proposal for the form of the interaction term was
\begin{eqnarray} \label{Lagrangian0}
{\mathcal{L}}=\sum_{\alpha=\pm}\sum_{l=1}^3 \frac{\lambda^{(\alpha,l)}}{M_{Pl}} \left[ \xi^{(\alpha,l)} \Xi^{(\alpha,l)}_{\mu \nu}+ \widetilde{\xi}^{(\alpha,l)} {\widetilde{\Xi}}^{(\alpha,l)}_{\mu \nu}\right] \bar\Psi \gamma^{\mu}\gamma^{\nu}\Psi,
\end{eqnarray}
where $\xi^{(\pm,l)}$ and $\widetilde{\xi}^{(\pm,l)}$ are some dimensionless coupling constants and the eigenforms are related via
\begin{equation} \label{Xitilde}
{\widetilde{\Xi}}^{(\pm,l)}_{\mu \nu}= {\epsilon_{\mu\nu}}^{\rho\sigma}\Xi^{(\pm,l)}_{\rho\sigma},
\end{equation}
with $\epsilon_{\mu\nu\rho\sigma}$ being the space-time volume $4$-form and the index $l$ running form $1$ to $3$. Note that the eigenforms $ \Xi^{(\pm,l)}_{\mu \nu}$ and ${\widetilde{\Xi}}^{(\pm,l)}_{\mu \nu}$ are both multiplied by the same eigenvalue. This is because there is an unavoidable degeneration of ${{(W_\pm)}_{\mu\nu}}^{\rho\sigma}$ caused by the symmetries of the Weyl tensor\footnote{We are considering only type I space-times according to Petrov classification where there are no further degeneracies \cite{Weyl matrix form}.}. Therefore, any linear combination of $ \Xi^{(\pm,l)}_{\mu \nu}$ and ${\widetilde{\Xi}}^{(\pm,l)}_{\mu \nu}$ is an eigenform of the corresponding operator and it is not clear which of these linear combinations must be used in the model. The proposed solution was to use those eigenforms satisfying
\begin{equation}
\label{eXX}\epsilon^{\mu\nu\rho\sigma}\Xi^{(\pm,l)}_{\mu \nu}\Xi^{(\pm,l)}_{\rho\sigma}=
\epsilon^{\mu\nu\rho\sigma}\widetilde{\Xi}^{(\pm,l)}_{\mu \nu}\widetilde{\Xi}^{(\pm,l)}_{\rho\sigma}=0.
\end{equation}
In addition, the norm of the eigenforms are related through
\begin{equation}
{\mathcal{G}} (\Xi^{(\pm,l)}, \Xi^{(\pm,l)})=-{\mathcal{G}}(\widetilde{\Xi}^{(\pm,l)}, \widetilde{\Xi}^{(\pm,l)}),
\end{equation}
so the positive normed eigenforms are denoted with a tilde and its norm is fixed to be $1$.

The fact that ${|\lambda^{(\pm,l)}|}^{1/2}/M_{Pl}$ is a dimensionless object is used to generalized the proposed term given in equation (\ref{Lagrangian0}) by the introduction of the parameters $s^{(\pm,l)}$ and $\widetilde{s}^{(\pm,l)}$. The most general coupling term is then\begin{small}
\begin{equation} \label{Lagrangian1}
{\mathcal{L}}=\sum_{\alpha,l}{|\lambda^{(\pm,l)}|}^{1/2}\left[ \xi^{(\alpha,l)} \left(\frac{|\lambda^{(\alpha,l)}|^{1/2}}{M_{Pl}}\right)^{s^{(\alpha,l)}} \Xi^{(\alpha,l)}_{\mu \nu}+ \widetilde{\xi}^{(\alpha,l)}\left(\frac{|\lambda^{(\alpha,l)}|^{1/2}}{M_{Pl}}\right)^{\widetilde{s}^{(\alpha,l)}} {\widetilde{\Xi}}^{(\alpha,l)}_{\mu \nu}\right] \bar\Psi \gamma^{\mu}\gamma^{\nu}\Psi,
\end{equation}\end{small}
where $s^{(\pm,l)}$ and $\widetilde{s}^{(\pm,l)}$ are required to be greater than $-1$ to ensure that in the limit of flat space-time this term vanishes.

In Ref. \cite{QGPwLSV} we noted that there was one further aspect that one would need to fix in order to have a truly unambiguous recipe for the desired Lagrangian term. The point is that if $\Xi_{\mu\nu}^{(\pm,l)}$ is an eigenform of ${{(W_{\pm})}_{\mu\nu}}^{\rho\sigma}$ satisfying the normalization condition and equation (\ref{eXX}), then $-\Xi_{\mu\nu}^{(\pm,l)}$ is another eigenform satisfying the same conditions. In that work we considered a simplistic solution and used the fact that the coupling constants $\xi^{(\pm,l)}$ and $\widetilde{\xi}^{(\pm,l)}$ can be used to absorb these signs and we argued that something like a `spontaneous symmetry breaking' mechanism might be at the source of the precise and full selection of the underlying space-time microstructure which we view as determining such ambiguities, and it was quite likely that the knowledge of the macroscopic space-time structure would not be enough to determine every detail of the way the space-time microstructure could become manifest at the phenomenological level. The specific aim of this work is to present a refined and fully unambiguous proposal for the possible manifestations of a fundamental granular micro-structure of space-time through its interaction with fermions and which does not violate Lorentz invariance.

\section{An unambiguous coupling term}

As we have discussed in the previous section and following \cite{NewQGP,QGPwLSV}, we seek to describe a possible influence of a space-time granularity on fermionic matter fields trough the construction of an object starting from the eigenvalues and eigenforms of some self-adjoint operators in (${\mathcal{S}},{\mathcal{G}}$) built from Weyl tensor, and which could be coupled to $\bar{\Psi}\gamma_{[\mu}\gamma_{\nu]}\Psi$. It is clear that the coupling term must be a Lorentz scalar and hermitian. In order to ensure 
that the eigenvalues and eigenvectors appearing in the Lagrangian are real we will work with a new pair of self-adjoint operators
\begin{eqnarray}
\label{W2+}{{({\mathcal{W}}_+)}_{\mu\nu}}^{\rho\sigma}&=&\frac{1}{2}\left({W_{\mu\nu}}^{\rho\sigma}+{W^\dagger_{\mu\nu}}^{\rho\sigma}\right),\\
\label{W2-}{{({\mathcal{W}}_-)}_{\mu\nu}}^{\rho\sigma}&=& \frac{1}{4}{\epsilon_{\mu\nu}}^{\alpha\beta}\left({W^\dagger_{\alpha\beta}}^{\rho\sigma}-{W_{\alpha\beta}}^{\rho\sigma}\right).
\end{eqnarray}
Moreover, and in contrast with our earlier proposals, we require that the Lagrangian density describing this coupling should be independent of the choice of sign of the eigenforms of the operators (\ref{W2+}-\ref{W2-}) avoiding any ambiguity in the proposal. This can be accomplished through the use the function $sign(x) $ which assigns to each real argument $x$ the value $1$, $-1$ or $0$ when $x$ is respectively positive, negative or zero. It is easy to see that for two $2$-forms $X_{\mu\nu}$ and $Y_{\mu\nu}$ the products $sign[{\mathcal{G}}(X,Y)]X_{\mu\nu} Y_{\rho\sigma}$ and $sign[{\mathcal{\epsilon}}(X,Y)]X_{\mu\nu} Y_{\rho\sigma}$ are independent of the signs of the $2$-forms. These terms can be combined with $\bar{\Psi}\gamma_{[\mu}\gamma_{\nu]}\Psi$
to form a non-trivial Lorentz scalar as follows:
\begin{equation}
\left\{ sign[{\mathcal{G}}(X,Y)]+\chi sign[{\mathcal{\epsilon}}(X,Y)]\right\} g^{\rho\sigma}X_{\rho [\mu}Y_{\nu]\sigma}\bar{\Psi}\gamma^{[\mu}\gamma^{\nu]}\Psi,
\end{equation}
where $\chi$ is some constant. This way we obtain a simple manner in which the eigenforms of ${{({\mathcal{W}}_\pm)}_{\mu\nu}}^{\rho\sigma}$ can be combined to form the type of unambiguous coupling term we are looking for.

Let us now see this construction in more detail. Each of the operators ${{({\mathcal{W}}_\pm)}_{\mu\nu}}^{\rho\sigma}$ has three different eigenvalues and three negative-normed eigenforms, which are respectively given by $\kappa^{(\pm,l)}$ and $\Theta^{(\pm,l)}_{\mu\nu}$, where we follow the same convention in the super-index as in the previous section. Hence, the most general coupling term containing these objects and having the required properties can be written as
\begin{eqnarray}\label{lagrangian2}
{\mathcal{L}}=& \bar{\Psi} \gamma^\mu \gamma^\nu \Psi g^{\rho\sigma}\sum_{\alpha,\beta=\pm}\sum_{l,m=1}^3\\
&\left\{\left( M^{(\alpha,\beta,l,m)} sign[{\mathcal{G}}(\Theta^{(\alpha,l)},\Theta^{(\beta,m)})]
+N^{(\alpha,\beta,l,m)} sign[\epsilon(\Theta^{(\alpha,l)},\Theta^{(\beta,m)})]
\right)\Theta^{(\alpha,l)}_{\rho[\mu}\Theta^{(\beta,m)}_{\nu]\sigma}\right. \nonumber\\ + &\left.\left(\widetilde{M}^{(\alpha,\beta,l,m)}sign[{\mathcal{G}}(\Theta^{(\alpha,l)},\widetilde{\Theta}^{(\beta,m)})]+\widetilde{N}^{(\alpha,\beta,l,m)}sign[\epsilon(\Theta^{(\alpha,l)},\widetilde{\Theta}^{(\beta,m)})]\right)\Theta^{(\alpha,l)}_{\rho[\mu}\widetilde{\Theta}^{(\beta,m)}_{\nu]\sigma}\right\},\nonumber
\end{eqnarray}
where
\begin{small}
\begin{eqnarray}
M^{(\alpha,\beta,l,m)}&=&\xi^{(\alpha,\beta,l,m)} |\kappa^{(\alpha,l)}|^{1/4}|\kappa^{(\beta,m)}|^{1/4} \left(\frac{|\kappa^{(\alpha,l)}|^{1/2}}{M_{Pl}}\right)^{c^{(\alpha,l)}}\left(\frac{|\kappa^{(\beta,m)}|^{1/2}}{M_{Pl}}\right)^{c^{(\beta,m)}},\\
N^{(\alpha,\beta,l,m)}&=&\chi^{(\alpha,\beta,l,m)} |\kappa^{(\alpha,l)}|^{1/4}|\kappa^{(\beta,m)}|^{1/4} \left(\frac{|\kappa^{(\alpha,l)}|^{1/2}}{M_{Pl}}\right)^{d^{(\alpha,l)}}\left(\frac{|\kappa^{(\beta,m)}|^{1/2}}{M_{Pl}}\right)^{d^{(\beta,m)}},\\
\widetilde{M}^{(\alpha,\beta,l,m)}&=&\widetilde{\xi}^{(\alpha,\beta,l,m)}|\kappa^{(\alpha,l)}|^{1/4}|\kappa^{(\beta,m)}|^{1/4}\left(\frac{|\kappa^{(\alpha,l)}|^{1/2}}{M_{Pl}}\right)^{\widetilde{c}^{(\alpha,l)}}\left(\frac{|\kappa^{(\beta,m)}|^{1/2}}{M_{Pl}}\right)^{\widetilde{c}^{(\beta,m)}},\\
\widetilde{N}^{(\alpha,\beta,l,m)}&=&\widetilde{\chi}^{(\alpha,\beta,l,m)}|\kappa^{(\alpha,l)}|^{1/4}|\kappa^{(\beta,m)}|^{1/4}\left(\frac{|\kappa^{(\alpha,l)}|^{1/2}}{M_{Pl}}\right)^{\widetilde{d}^{(\alpha,l)}}\left(\frac{|\kappa^{(\beta,m)}|^{1/2}}{M_{Pl}}\right)^{\widetilde{d}^{(\beta,m)}},
\end{eqnarray}\end{small}
and in these last expressions $\xi^{(\alpha,\beta,l,m)}$, $\widetilde{\xi}^{(\alpha,\beta,l,m)}$, $\chi^{(\alpha,\beta,l,m)}$, $\widetilde{\chi}^{(\alpha,\beta,l,m)}$, $c^{(\alpha,l)}$, $\widetilde{c}^{(\alpha,l)}$, $d^{(\alpha,l)}$ and $\widetilde{d}^{(\alpha,l)}$ are the free dimensionless parameters of the model with the restriction that
\begin{equation}
c^{(\alpha,l)},\widetilde{c}^{(\alpha,l)},d^{(\alpha,l)},\widetilde{d}^{(\alpha,l)}>-1/2,
\end{equation}
to ensure that there is no divergence if an eigenvalue is zero (as in the case of flat space-time). The $\widetilde{\Theta}^{(\pm,l)}_{\mu\nu}$ appearing in equation (\ref{lagrangian2}) are the positive-normed eigenforms and they are related to $\Theta^{(\pm,l)}_{\mu\nu}$ by a relation which is analogous to (\ref{Xitilde}). In addition, the eigenforms are normalized to $\pm 1$ and the negative normed are chosen to be such that
\begin{equation}
\epsilon^{\mu\nu\rho\sigma}\Theta^{(\pm,l)}_{\mu\nu}\Theta^{(\pm,l)}_{\rho\sigma}=0.
\end{equation}

It is important to note that it is not possible to fix the ambiguities using the terms containing the same eigenform twice, however, it is easy to see that such terms vanish automatically. On the other hand, the fact that $\epsilon_{\mu\nu\rho\sigma}\epsilon^{\mu\nu\kappa\lambda}=-4\delta_{[\rho}^\kappa\delta_{\sigma]}^\lambda$ and $\epsilon_{\mu\nu\rho\sigma}\epsilon^{\mu\tau\kappa\lambda}=-6\delta_{[\nu}^\tau\delta_{\rho}^\kappa\delta_{\sigma]}^\lambda$ lead to the following identities
\begin{eqnarray}
X_{\mu [\nu}{\widetilde{Y}_{\rho ]}}^\mu&=&\widetilde{X}_{\mu [ \nu}{Y_{\rho ]}}^\mu,\\
\widetilde{X}_{\mu [ \nu}{\widetilde{Y}_{\rho ]}}^\mu&=&-4 X_{\mu [ \nu}{Y_{\rho ]}}^\mu,
\end{eqnarray}
which ensure that we do not need to consider more terms in equation (\ref{lagrangian2}). In addition, two eigenforms with $\alpha=\beta$ but $l\neq m$ are $\mathcal{G}$-orthogonal given that they are eigenforms of a $\mathcal{G}$-self-adjoint operator. Therefore, the terms containing $sign[{\mathcal{G}}(\Theta^{(\alpha,l)},\Theta^{(\alpha,m)})]$ do not appear in the Lagrangian. For the same reason, and using $\epsilon_{\mu\nu\kappa\lambda}  {\epsilon^{\kappa\lambda}}_{\rho \sigma}={-\mathcal{G}}_{\mu\nu\rho\sigma}$, it is easy to see that $\epsilon(\Theta^{(\alpha,l)},\widetilde{\Theta}^{(\alpha,m)}) = 0$ which also reduces the terms in the Lagrangian. In the following section we will obtain the corresponding effective low energy Hamiltonian which can be used in studying phenomenological implications of this model.

\section{The effective low energy Hamiltonian}

We consider here situations in which the linearized gravity approximation is justified as it will be in all conceivable experiments in the solar system and in a laboratory. We are interested in the effects of the gravitational environment on test particles and therefore, when describing how the local matter distribution determines the gravitational environment, we only need to consider the standard Einstein equations. This is justified because on the one hand the corrections on Einstein equations that would arise from the new couplings are of higher order in the gravitational constant $G$, but even more importantly because the ordinary bulk matter in a laboratory is not polarized, and given the spin dependence of the new matter-gravity couplings it is clear that it will not act as a source. In the regime of interest it is enough to consider the lowest order perturbative analysis, so we write
\begin{equation}
g_{\mu\nu}=\eta_{\mu\nu}+ \gamma_{\mu\nu}, 
\end{equation}
where $\eta_{\mu\nu}$ is a flat space-time metric\footnote{We are using the convention where the space-time metric has signature $(-,+,+,+)$.} and $\gamma_{\mu\nu}$ is a small perturbation. We take the Minkowskian coordinates associated with $\eta_{\mu\nu}$ as approximately identified with the laboratory measured coordinates $t, \vec{x} $. After fixing the gauge in the standard fashion, namely by imposing $\partial^\mu \bar{\gamma}_{\mu\nu}=0$ where $\bar{\gamma}_{\mu\nu}=\gamma_{\mu\nu}-\frac{1}{2}\eta_{\mu\nu}\eta^{\rho\sigma}\gamma_{\rho\sigma}$ (see \cite{Wald}), we focus on the structure of the Weyl tensor. As indicated, we only consider the situations in which the sources vanish at the points to be probed experimentally, thus
\begin{equation} \label{weyllin2} 
{W_{\mu\nu}}^{\rho\sigma}={R_{\mu\nu}}^{\rho\sigma}=-2\partial^{[\rho} \partial_{[\mu} \bar{\gamma}_{\nu]}^{\sigma]}+\delta^{[\rho}_{[\mu} \partial_{\nu]} \partial^{\sigma]} \bar{\gamma},
\end{equation}
where $\bar{\gamma} =\eta^{\mu\nu}\bar{\gamma}_{\mu\nu}$. The linearized Einstein equations are 
\begin{equation} \label{Helmholtz}
\partial_\rho \partial^\rho \bar{\gamma}_{\mu\nu}=-16\pi G T_{\mu\nu},
\end{equation}
and the appropriate solution is obtained in the standard way by means of the retarded Green function
\begin{equation} \label{Green}
\bar{ \gamma}_{\mu\nu}=4G\int T_{\mu\nu}(x')\frac{\delta(t'-t_r)}{R}d^4x',
\end{equation}
where $R= |\vec{x}-\vec{x}'|$ and $t_r = t- R/c$ is the retarded time. Note that in this section we are re-inserting the factors $1/c$ because we will be interested in the dominant order.

To first order in $1/c$, we write the components of the energy momentum tensor in the laboratory frame as 
\begin{equation}
T_{\mu\nu}=\delta_\mu^0 \delta_\nu^0\rho +\frac{1}{c}\left(\delta_\mu^0\delta_\nu^k+\delta_\mu^k\delta_\nu^0\right)p_k,
\end{equation}
where $\rho$ is the matter density and $p_k/c$ represent the $k$-th component of the momentum density. Observe that the conservation equations in the linearized regime, $\partial^\mu T_{\mu\nu}=0$, imply
\begin{equation}
\label{cont}\dot{\rho}=\nabla\cdot\vec{p},
\end{equation}
where the dot represent a derivative with respect to the time coordinate. From equation (\ref{Green}) and neglecting the terms $1/c^2$ it can be seen that
\begin{eqnarray}
\bar{ \gamma}_{\mu\nu}&=&4G\left(\delta_\mu^0 \delta_\nu^0 \int \frac{\rho(\vec{x}', t_r)}{R}d^3x'+\frac{2}{c}\delta_{(\mu}^0\delta_{\nu)}^k\int \frac{p_k(\vec{x}', t_r)}{R}d^3x'\right)\nonumber\\
&=& 4G\left[\delta_\mu^0 \delta_\nu^0\left(\int \frac{\rho(\vec{x}', t)}{R}d^3x'-\frac{1}{c} \int\dot{\rho}(\vec{x}', t)d^3x'\right)+\frac{2}{c}\delta_{(\mu}^0\delta_{\nu)}^k\int \frac{p_k(\vec{x}', t)}{R}d^3x'\right]\nonumber\\
&=& 4G\left[\delta_\mu^0 \delta_\nu^0\int \frac{\rho(\vec{x}', t)}{R}d^3x'+\frac{2}{c}\delta_{(\mu}^0\delta_{\nu)}^k\int \frac{p_k(\vec{x}', t)}{R}d^3x'\right],
\end{eqnarray} 
where in the second step we do a Taylor expansion around $ t_r = t$ and in the last step we have used equation (\ref{cont}), the divergence theorem and the fact that the gravitational sources are taken to be compact. Next we define
\begin{eqnarray}
\label{PhiNewtoniano}\Phi_N&=& G \int \frac{\rho(\vec{x}', t)}{R}d^3x' ,\\
\label{Pi}\Pi^k&=& G \int \frac{p^k(\vec{x}', t)}{R}d^3x',
\end{eqnarray}
where $\Phi_N$ corresponds to the standard Newtonian potential due to the matter source. We then can write
\begin{equation}\label{gammabar}
\bar{ \gamma}_{\mu\nu}=4\left[\delta_\mu^0 \delta_\nu^0\Phi_N+\frac{2}{c}\delta_{(\mu}^0\delta_{\nu)}^k\Pi_k \right],
\end{equation}
and using equation (\ref{weyllin2}) we get
\begin{equation} \label{weyllin3} 
{W_{\mu\nu}}^{\rho\sigma}=-4 \left(2\delta_{[\mu}^0 \eta^{0[\rho}+\delta_{[\mu}^{[\rho}\right)\partial_{\nu]} \partial^{\sigma]}\Phi_N -\frac{8}{c}\left(\delta_{[\mu}^{0}\eta^{k[\rho}+\delta_{[\mu}^{k}\eta^{0[\rho}\right)\partial_{\nu]}\partial^{\sigma]}\Pi_k.
\end{equation}

At this point is useful to introduce a new notation where the capital letters $A$, $B$, $C$ and $D$ represent antisymmetric pairs of space-time indexes and which are numerated with roman numerals with the convention $I=01$, $II=02$, $III=03$, $IV=23$, $V=31$ and $VI=12$. With this notation any $(2,2)$ tensor ${T_{\mu\nu}}^{\rho\sigma}$ which is antisymmetric in its two pairs of indexes can be expressed as a $6 \times 6$ matrix given by
\begin{eqnarray}
{T_A}^B \equiv \left(\begin{array}{cccc}
{T_I}^I & {T_I}^{II} &\cdots & {T_I}^{VI} \\
{T_{II}}^I & {T_{II}}^{II} & \cdots & {T_{II}}^{VI} \\
& & & \\
\vdots & \vdots & \ddots & \vdots \\
& & & \\
{T_{VI}}^I & {T_{VI}}^{II} & \cdots & {T_{VI}}^{VI} 
\end{array} \right)= \left(\begin{array}{cccccc}
&&&&&\\
&\mathbf{K}&&&\mathbf{L}&\\
&&&&&\\
&&&&&\\
&\mathbf{M}&&&\mathbf{N}&\\
&&&&&
\end{array} \right),
\end{eqnarray}
where boldface capital letters represent $3\times 3$ matrices. In particular, due to its symmetries, the Weyl tensor has the generic form \cite{Weyl matrix form} 
\begin{eqnarray} \label{Weylgral}
{W_A}^B= \left(\begin{array}{cccccc}
&&&&&\\
&\mathbf{ A}&&&\mathbf{ B}&\\
&&&&&\\
&&&&&\\
&-\mathbf{B}&&&\mathbf{ A}&\\
&&&&&
\end{array} \right),
\end{eqnarray}
where $\mathbf{A}$ and $\mathbf{B}$ are $3 \times 3$ real traceless symmetric matrices. From the upper-left side of the matrix in equation (\ref{Weylgral}) it is possible to read that the components of $\mathbf{A}$ are
\begin{equation}\label{A}
{A_i}^j={W_{0i}}^{0j}= \partial_{i} \partial^{j}\Phi_N,
\end{equation}
where the terms suppressed by $c^{-2}$ are omitted. Note that $\mathbf{A}$ is symmetric and it can be seen, using the Poisson equation and the fact that the matter density of the source at the points of interest vanishes, that is traceless as required. With a similar method it is possible to show that the components of $\mathbf{B}$ satisfy
\begin{eqnarray}
{B_i}^j&=&\frac{1}{4} ({\epsilon^{j}}_{mn}{W_{0i}}^{mn}-{\epsilon_{i}}^{mn}{W_{mn}}^{0j})\nonumber\\
&=&\frac{1}{2c} \left[{\epsilon^{j}}_{mn}(\delta_i^{[m}\partial^{n]}\dot{\Phi}_N+2\partial_i\partial^{[m}\Pi^{n]})-{\epsilon_{i}}^{mn}(-\delta_{[m}^j\partial_{n]}\dot{\Phi}_N-2\partial^{j}\partial_{[m}\Pi_{n]})\right]\nonumber\\
&=&\frac{1}{c}\left[\partial_i{(\nabla\times\vec{\Pi})}^j+\partial^j{(\nabla\times\vec{\Pi})}_i\right],\label{B}
\end{eqnarray}
where the first identity can be verified by comparison with equation (\ref{Weylgral}) and using the traceless property of Weyl tensor, and in the second step we use
\begin{eqnarray}
{W_{0i}}^{mn}&=&\frac{1}{c}\left(2\delta_i^{[m}\partial^{n]}\dot{\Phi}_N+4\partial_i\partial^{[m}\Pi^{n]}\right),\\
{W_{mn}}^{0j}&=&\frac{1}{c}\left(-2\delta_{[m}^j\partial_{n]}\dot{\Phi}_N-4\partial^{j}\partial_{[m}\Pi_{n]}\right),
\end{eqnarray}
that are obtained from equation (\ref{weyllin3}). As is required by the properties of the Weyl tensor, $\mathbf{B}$ is also symmetric and its trace vanishes because it is a divergence of a rotational. Furthermore, by using the equations (\ref{PhiNewtoniano}-\ref{Pi}) we can re-express the components of $\mathbf{A}$ and $\mathbf{B}$ as
\begin{eqnarray}
{A_i}^j&=& G\int \frac{3(x_i-x_i')(x^j-x'^j)-R^2\delta_i^j }{R^5}\rho(\vec{x}', t)d^3x',\\
{B_i}^j&=&\frac{G}{c}\int \frac{(x_i-x_i') L^j+(x^j-x'^j) L_i}{R^5}d^3x',
\end{eqnarray}
where $L_i=\epsilon_{ijk}(x^j-x'^j)p^k$.

In this notation and using a normal Riemannian coordinate system associated with the point where the probe is located, the supermetric and the volume element in that point can be expressed as
\begin{eqnarray}
{\mathcal{G}}_{AB}=\left(\begin{array}{cccccc}
&&&&&\\
&-\mathbf{1}&&&\mathbf{ 0}&\\
&&&&&\\
&&&&&\\
&\mathbf{0}&&&\mathbf{1}&\\
&&&&&
\end{array} \right),
\ \ \ \ \ 
{\epsilon_A}^B = \left(\begin{array}{cccccc}
&&&&&\\
&\mathbf{ 0}&&&\mathbf{ 1}&\\
&&&&&\\
&&&&&\\
&-\mathbf{1}&&&\mathbf{0}&\\
&&&&&
\end{array} \right),
\end{eqnarray}
where $\mathbf{ 0}$ and $\mathbf{1}$ respectively stand for the $3 \times 3$ zero and identity matrices. Furthermore, the hermitian operators have the matricial forms
\begin{eqnarray}
{{({\mathcal{W}}_+)}_{A}}^{B}= \left(\begin{array}{cccccc}
&&&&&\\
&\mathbf{ A}&&&\mathbf{ 0}&\\
&&&&&\\
&&&&&\\
&\mathbf{0}&&&\mathbf{ A}&\\
&&&&&
\end{array} \right),\ \ \ \ \ 
{{({\mathcal{W}}_-)}_{A}}^{B}= \left(\begin{array}{cccccc}
&&&&&\\
&\mathbf{ B}&&& \mathbf{0}&\\
&&&&&\\
&&&&&\\
&\mathbf{0}&&&\mathbf{B}&\\
&&&&&
\end{array} \right).
\end{eqnarray}
Hence, in order to obtain the eigenvalues and eigenforms of ${({\mathcal{W}}_\pm)_A}^B$, we first solve the eigenvalue problem for $\mathbf{A}$ and $\mathbf{B}$. Assuming that the vectors $\vec{a}^{(l)}$, $\vec{b}^{(l)}$ and the coefficients $\alpha^{(l)}$, $\beta^{(l)}$ such that
\begin{eqnarray}
{A_i}^ja^{(l)}_j=\alpha^{(l)} a^{(l)}_i,\ \ \ \ \ 
{B_i}^jb^{(l)}_j=\beta^{(l)} b^{(l)}_i,
\end{eqnarray}
are known (where, we recall, $l=1,2,3$), then, the eigenvalues of ${({\mathcal{W}}_\pm)_A}^B$ are 
\begin{eqnarray}
\kappa^{(+,l)}=\alpha^{(l)},\ \ \ \ \ 
\kappa^{(-,l)}=\beta^{(l)},
\end{eqnarray}
and the negative-normed eigenforms, which in this notation are written as $6$-column vectors, take the form
\begin{eqnarray}\label{Theta+-}
\Theta_A^{(+,l)}= \left(\begin{array}{c}\vec{a}^{(l)} \\ \\ \vec{0}\end{array} \right),
\ \ \ \ \ 
\Theta_A^{(-,l)}= \left(\begin{array}{c}\vec{b}^{(l)} \\ \\ \vec{0}\end{array} \right).
\end{eqnarray}
It is important to note that $\mathbf{A}$ and $\mathbf{B}$ are symmetric and real, therefore their eigenvalues and the components of the corresponding eigenvectors are real. We also note that, when the eigenvectors of $\mathbf{A}$ and $\mathbf{B}$ are taken to form a orthonormal triad, the corresponding eigenforms (\ref{Theta+-}) are automatically normalized to $-1$. Furthermore, these eigenforms satisfy condition (\ref{eXX}) and thus, they are the first set of eigenforms that are required by the model, while the positive-normed eigenforms, which are obtained using the analog of equation (\ref{Xitilde}), are
\begin{eqnarray}
\widetilde{\Theta}_A^{(+,l)}= \left(\begin{array}{c}\vec{0} \\ \\ -\vec{a}^{(l)}\end{array} \right),
\ \ \ \ \ 
\widetilde{\Theta}_A^{(-,l)}= \left(\begin{array}{c}\vec{0} \\ \\ -\vec{b}^{(l)}\end{array} \right).
\end{eqnarray}

We should point out that $\alpha^{(l)}$ and $\beta^{(l)}$ are related to the complex eigenvalues of the full Weyl tensor, $\omega^{(l)}$, through $\omega^{(l)}=\alpha^{(l)}+ i \beta^{(l)}$. In addition, $\omega^{(l)}$ satisfy the simple relations \cite{Eigenvalues}
\begin{eqnarray}
\omega^{(1)}+\omega^{(2)}+\omega^{(3)}&=&0,\\
\omega^{(1)} \omega^{(2)} + \omega^{(1)} \omega^{(3)} + \omega^{(2)}\omega^{(3)} &=&-\frac{1}{32}{W^{+}}_{\mu\nu\rho\sigma} {W^{+}}^{\mu\nu\rho\sigma},\\
\omega^{(1)}\omega^{(2)}\omega^{(3)} &=& -\frac{1}{192}\left({{W^{+}}_{\mu\nu}}^{\rho\sigma}{{W^{+}}_{\rho\sigma}}^{\alpha\beta}{{W^{+}}_{\alpha \beta}}^{\mu\nu}\right),
\end{eqnarray}
where ${W^{+}}_{\mu\nu\rho\sigma}= W_{\mu\nu\rho\sigma}+i{}^*W_{\mu\nu\rho\sigma}$ and ${}^*W_{\mu\nu\rho\sigma}$ denotes Weyl dual. These three complex equations allow us to express the quantities $\alpha^{(l)}$ and $\beta^{(l)}$ in terms of the Weyl tensor indicating that they are not as `exotic' as they might seem at first sight.

In order to obtain an effective low energy Hamiltonian, which could be used in phenomenological analysis, we note that the interaction term has the same form of the $-1/2 H_{\mu \nu} \bar\Psi \gamma^{[\mu} \gamma^{\nu]}\Psi$ term considered in the Standard Model Extension (SME) \cite{SME}. Comparing with equation (\ref{lagrangian2}), we can connect the two formalisms by identifying $H_{\mu \nu}$ with\begin{small}
\begin{eqnarray} \label{Hmunu}
H_{\mu \nu}=&-2 g^{\rho\sigma}\sum_{\alpha,\beta=\pm}\sum_{l,m=1}^3\\
&\left\{\left( M^{(\alpha,\beta,l,m)} sign[{\mathcal{G}}(\Theta^{(\alpha,l)},\Theta^{(\beta,m)})]
+N^{(\alpha,\beta,l,m)} sign[\epsilon(\Theta^{(\alpha,l)},\Theta^{(\beta,m)})]
\right)\Theta^{(\alpha,l)}_{\rho[\mu}\Theta^{(\beta,m)}_{\nu]\sigma}\right. \nonumber\\ + &\left.\left(\widetilde{M}^{(\alpha,\beta,l,m)}sign[{\mathcal{G}}(\Theta^{(\alpha,l)},\widetilde{\Theta}^{(\beta,m)})]+\widetilde{N}^{(\alpha,\beta,l,m)}sign[\epsilon(\Theta^{(\alpha,l)},\widetilde{\Theta}^{(\beta,m)})]\right)\Theta^{(\alpha,l)}_{\rho[\mu}\widetilde{\Theta}^{(\beta,m)}_{\nu]\sigma}\right\}\nonumber.
\end{eqnarray}\end{small}
The relevant contributions to the non-relativistic Hamiltonian can be directly read of from the formulation of Ref. \cite{NRH} to be
\begin{equation} \label{HNR0}
{\mathcal{H}}_{NR}= \epsilon^{ijk} \left[\frac{1}{2}\left(\sigma_i + \left(\vec{\sigma}\cdot \frac{\vec{P}}{M} \right) \frac{P_i}{M} \right) H_{jk} + \left(1- \frac{1}{2}\frac{P^2}{M^2} \right) \frac{P_i}{M} \sigma_j H_{0k} \right],
\end{equation}
where $\vec{P}$ and $M$ are respectively the momentum and mass of the test particle and $\sigma_i$ are the Pauli matrices. It should be noted that in contrast with Ref. \cite{NRH} where the $H_{\mu \nu}$ are `fixed', and in contrast with the curved space-time version of the SME \cite{SME2} where the objects like $H_{\mu\nu}$ are new fields (and thus space-time dependent), in the proposal we are presenting the object that plays the corresponding role is completely determined by the surrounding gravitational environment. However, we must warn the reader that the fact that the objects $H_{\mu \nu}$ are in principle space-time dependent has not been taken into account in the derivation of the effective Hamiltonian of equation (\ref{HNR0}), and thus, its validity is restricted to situations where its components in the Riemann normal coordinates are essentially constant in the region where the probing particle is localized.

From equation (\ref{Hmunu}) and using that in the case under consideration $\Theta_{jk}^{(\pm,l)}=\widetilde{\Theta}_{0i}^{(\pm,l)}=0$ as well as the fact that some contractions of the eigenforms with ${\mathcal{G}}_{\mu\nu\rho\sigma}$ and $\epsilon_{\mu\nu\rho\sigma}$ vanish, it can be shown that
\begin{eqnarray}
H_{0i}=&g^{jk}\sum_{\alpha,\beta=\pm} \sum_{l,m=1}^3\widetilde{N}^{(\alpha,\beta,l,m)}sign[\epsilon(\Theta^{(\alpha,l)},\widetilde{\Theta}^{(\beta,m)})]\Theta^{(\alpha,l)}_{0j}\widetilde{\Theta}^{(\beta,m)}_{ik}\nonumber\\
=&\sum_{l,m=1}^3(\widetilde{N}^{(+,-,l,m)}-\widetilde{N}^{(-,+,m,l)})sign[\vec{a}^{(l)}\cdot \vec{b}^{(m)}] (\vec{a}^{(l)}\times \vec{b}^{(m)})_i,\label{H0i}
\end{eqnarray}
and similarly,
\begin{eqnarray}
\frac{1}{2}{\epsilon_i}^{jk}H_{jk}
=&{\epsilon_i}^{jk} \sum_{\alpha,\beta=\pm}\sum_{l,m=1}^3
M^{(\alpha,\beta,l,m)} sign[{\mathcal{G}}(\Theta^{(\alpha,l)},\Theta^{(\beta,m)})]
\Theta^{(\alpha,l)}_{0[j}\Theta^{(\beta,m)}_{k]0}\nonumber\\
=&\sum_{l,m=1}^3 \left(M^{(+,-,l,m)}-M^{(-,+,m,l)}\right) sign[\vec{a}^{(l)}\cdot\vec{b}^{(m)}]
(\vec{a}^{(l)}\times\vec{b}^{(m)})_i.\label{Hjk}
\end{eqnarray}
With expressions (\ref{H0i}-\ref{Hjk}) at hand we can write
\begin{eqnarray}\label{HNR}
{\mathcal{H}}_{NR}=&\sum_{l,m=1}^3 sign[\vec{a}^{(l)}\cdot \vec{b}^{(m)}] (\vec{a}^{(l)}\times \vec{b}^{(m)})_k \nonumber\\ & \left[\left(\sigma^k + \left(\vec{\sigma}\cdot \frac{\vec{P}}{M} \right) \frac{P^k}{M} \right)  \left(M^{(+,-,l,m)}-M^{(-,+,m,l)}\right)\right.\nonumber\\
&\left. + \left(1- \frac{1}{2}\frac{P^2}{M^2} \right) \frac{(\vec{P}\times \vec{\sigma})^k}{M} (\widetilde{N}^{(+,-,l,m)}-\widetilde{N}^{(-,+,m,l)}) \right],
\end{eqnarray}
which is the phenomenological Hamiltonian that can be compared with experimental measurements in order to look for traces of the interaction of a Lorentz invariant granularity of space-time and fermions and to get bounds on the free parameters of the model. Observe that in a reference frame where the probing particle is at rest, the effective Hamiltonian takes the simpler form
\begin{eqnarray}\label{HNRs}
{\mathcal{H}}_{NR}=\sum_{l,m=1}^3 \left(M^{(+,-,l,m)}-M^{(-,+,m,l)}\right) sign[\vec{a}^{(l)}\cdot \vec{b}^{(m)}] \vec{\sigma}\cdot\vec{a}^{(l)}\times \vec{b}^{(m)} .\end{eqnarray}
It should also be stressed that eventhough that the model presented in this work is completely covariant, the above expressions for $H_{0i}$ and $H_{jk}$ do not seem to be  covariantly related. The reason for this is that we are neglecting the $c^{-2}$ terms and the Lorentz transformation that mixes $H_{0i}$ with $H_{jk}$ involves a factor $c^{-1}$.

In Ref. \cite{BoundsH0i} we find the bounds that have been obtained for the quantities  $H_{0i}$ that appear in the SME trough the analysis of a Hughes-Drever-like experiment which monitored certain spectral lines and their variations with a period of a year (taking it to correspond to the modulation of the field components in the Earth's lab when assuming that the the components of the fields remain fixed in the CMB frame). The bounds obtained can be characterized in a simple way by  $H_{0i}\leq10^{-27}\ \mbox{GeV}$. When considering the implications of these bounds for our model we first concluded that they would not be relevant simply because, in contrast with the SME setting, our fields in the lab would be determined by the local gravitational environment  which would not involve a modulation with a period of one year. However, after a little thought we realized that the tidal forces exerted by the Sun over the objects on Earth would vary with precisely a one year period due to the ellipticity of the Earth's orbit.  Thus, we calculated the effective $H_{0i}$ corresponding to such tidal forces according to our model and its modulation due to the the shape of the Earth's orbit. Considering the differences in $H_{0i}$ during the perihelion and the aphelion and comparing with the bounds from Ref. \cite{BoundsH0i} we obtained the following constraints for some of the parameters of our model:
\begin{eqnarray}
e^{-115 \widetilde{d}^{(+,1)}-120 \widetilde{d}^{(-,2)}} \left|\widetilde{\chi}^{(+,-,1,2)}-\widetilde{\chi}^{(-,+,2,1)}\right|&\leq& 10^{6},\\
e^{-115 \widetilde{d}^{(+,3)}-120 \widetilde{d}^{(-,2)}} \left|\widetilde{\chi}^{(+,-,3,2)}-\widetilde{\chi}^{(-,+,2,3)}\right|&\leq& 10^{6},\\
e^{-115 \widetilde{d}^{(+,3)}-120 \widetilde{d}^{(-,3)}} \left|\widetilde{\chi}^{(+,-,3,3)}-\widetilde{\chi}^{(-,+,3,3)}\right|&\leq& 10^{6},\\
e^{-115 \widetilde{d}^{(+,1)}-120 \widetilde{d}^{(-,3)}} \left|\widetilde{\chi}^{(+,-,1,3)}-\widetilde{\chi}^{(-,+,3,1)}\right|&\leq& 10^{6}.
\end{eqnarray}
The above are the first bounds obtained for the free parameters of the model but we are certain that stronger bounds can be set by, for example, analyzing the effects of the tidal forces due to the Moon.

In order to summarize we give an outline of the procedure that, according to this work, has to be followed to search for the proposed effects. First, given the gravitational sources of a particular experimental setting, its Newtonian Potential $\Phi_N$ and the vector $\vec{\Pi}$ have to be calculated through equations (\ref{PhiNewtoniano}-\ref{Pi}). Then, the matrices $\mathbf{A}$ and $\mathbf{B}$ presented in equations (\ref{A}-\ref{B}) have to be diagonalized and finally, its eigenvalues ($\alpha^{(l)}$, $\beta^{(l)}$) and eigenvectors ($\vec{a}^{(l)}$, $\vec{b}^{(l)}$) have to be plugged into the low energy Hamiltonian (\ref{HNR}).

\section{Conclusions}

We have presented a model where fermionic matter fields `feel' a non-trivial micro-structure of space-time. This micro-structure respects Lorentz symmetry, as is strongly suggested by experimental and theoretical results, but besides this, there is no need of specifying its structure. This model is an improvement of our previous work \cite{QGPwLSV} because in this case there are no undetermined signs. The main point of this manuscript is that it is possible to search for the effects that a Lorentz respectful micro-structure of space-time may induce in `quantum' particles.

Regarding the experimental outlook we note that the polarization dependence of the couplings, which can be inferred by the presence of the Pauli matrices in equation (\ref{HNR}), imply that in any relevant experiment, the test fermions must be polarized and in fact, one would need to measure the energy difference between the two polarizations. This energy difference is what should be compared with the difference of the eigenvalues of the operator (\ref{HNR}) to look for the proposed effect or to get bounds on the free parameters of the model. This requirement tends to make the conceivable experiments very difficult due to the standard magnetic couplings of polarized matter, although we must remark that the polarized matter probe designed to avoid those effects has been built \cite{Adelberger}, and thus, we can hope that actual tests of these ideas can be carried out in the near future. One aspect of the present proposal that distinguishes it at the phenomenological level from
the previous ones is the necessity of having a gravitational environment where both, tidal Newtonian forces (characterized by the matrix $\mathbf{A}$) and general relativistic dragging of frames (characterized by the matrix $\mathbf{B}$) are present to a non-negligible degree (in comparison with the sensitivity of the experiment). We refer the reader that is interested in a deeper discussion of the broader spectrum of experiments where these effects can be sough to Ref. \cite{QGPwLSV}.

Finally, we should briefly mention that in this proposal the equivalence principle appears to be violated at first sight because gravity (\textit{i.e.}, the effect of space-time curvature) is not fully eliminated in a free falling system. However, a deeper inspection reveals that what is being consider here is related to the fact that the probes of the space-time structure are, at the quantum level, extended objects and that, as such, could be sensitive to the departure from exact flatness of space-time. Moreover, the magnitude of this effects could be enhanced relative to the naive expectations by a putative granularity of space-time. In fact, we should point out that, while the direct extrapolation of general relativity to the microscopic world would indicate that there is nothing expected to be found in this search, it is rather unclear how would one justify exactly the naive expectations on anything like a fundamental principle such as the equivalence principle \cite{Unspekables}. This is because, at this time, there is no concrete proposal for a formulation of the equivalence principle that can be applied in general to quantum systems (see \cite{Lamrzhal} for an intriguing proposal) essentially because this is a local principle and quantum objects are never precisely localized.

\section*{Acknowledgments}

We wish to tank E.G. Adelberger and V.A. Kostelecky for useful comments and to M. Necht for bringing to our attention reference \cite{Eigenvalues}. This work was presented at the Symposium on Mathematical Physics held at Torun on June 2008 thanks to the PAEP-UNAM 2008 program and its development was partially supported by DGAPA-UNAM project IN119808.

\end{document}